\documentclass{article}
\usepackage{amsmath,amssymb,amsfonts,amsthm}
\usepackage{authblk}
\usepackage{graphicx}
\usepackage{dcolumn}
\usepackage{bm}
\usepackage{hyperref}
\usepackage{color}
\usepackage{comment}
\usepackage[utf8]{inputenc}
\usepackage[english]{babel}
\usepackage{dsfont}
\usepackage{appendix}
\usepackage{algorithm}

\title{From Spin Glass Theory to Optimization in High
Dimensions and Modern Artificial Neural Networks%
\thanks{Revised author version of Chapter 13 in
\textit{From Quantum Fields to Spin Glasses:
Seminal Contributions of Giorgio Parisi to Theoretical Physics},
edited by R.~Aloisio et al.,
Cambridge University Press (2026), pp.~273--302.
\url{https://doi.org/10.1017/9781009709620}}}

\author[1]{Riccardo Zecchina}
\author[1]{Gianmarco Perrupato}
\affil[1]{Department of Computing Sciences, Bocconi University, 20136 Milano, Italy}

\date{}

\begin{document}

\maketitle

\begin{abstract}

This chapter pays tribute to Giorgio Parisi by reviewing the influence of his work on spin glasses and replica symmetry breaking (RSB) on combinatorial optimization and artificial neural networks. We first introduce replica and cavity methods through the matching and traveling salesman problems, before discussing random satisfiability and the development of survey propagation. We then examine applications to neural networks, emphasizing the connections between the geometry of solution spaces, computational hardness, and learning. Particular attention is given to rare regions of high local entropy and their role in algorithmic accessibility. Examples involving overparameterized models and asymmetric recurrent networks illustrate how these ideas connect constrained optimization and inference with concepts developed in spin-glass theory.

\end{abstract}

\section{Introduction}

\begin{figure}
    \centering
    \includegraphics[width=0.8\textwidth]{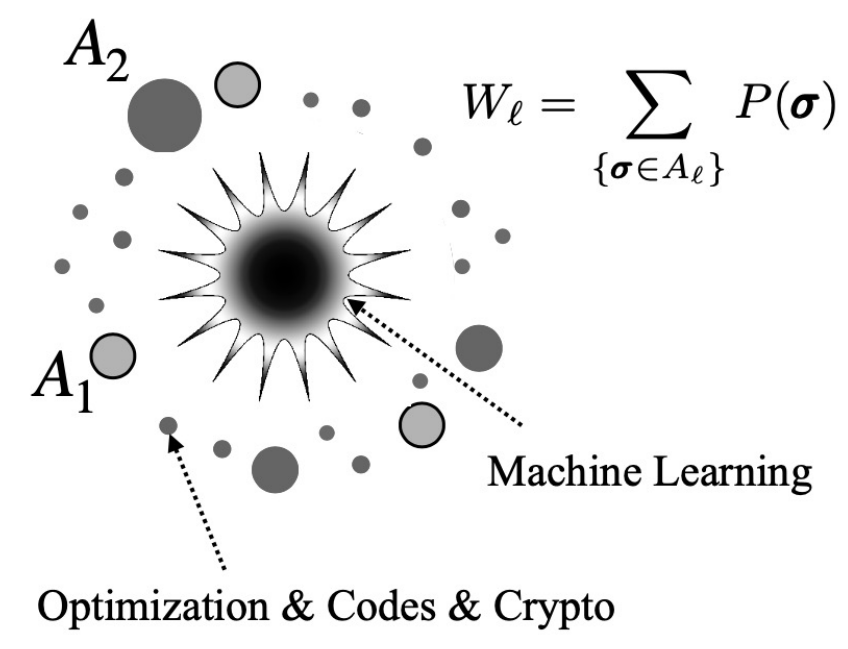}
    \caption{A pictorial representation of  different aspects of RSB which play a computational role in combinatorial optimization and learning.}
    \label{fig:scheme}
\end{figure}

The puzzling magnetic properties observed in spin glasses (SG) experiments \cite{mydosh1993spin} inspired a formidable intellectual challenge, leading to an exchange of paradigms  between statistical physics, optimization theory and artificial intelligence. A pivotal milestone in this scientific journey was reached at the end of the 70s by Giorgio Parisi, who developed the celebrated ``Replica Symmetry Breaking theory'' to solve the Sherrington-Kirkpatrick (SK) model \cite{kirkpatrick1978infinite} of SGs. The aim of this chapter is to discuss some examples of apparently very diverse non-convex optimization problems that fall within the unified geometrical framework evoked by replica symmetry breaking (RSB).
The objectives of different optimization problems are linked to distinct aspects of RSB. Information theoretic issues generally pertain to the characteristics of the dominant states. The difficulty of optimization is associated with the freezing phenomena in sub-dominant states, while the process of learning relates to entropic maximization and the presence of rare regions with high local entropy.  A pictorial view is given in Fig.~\ref{fig:scheme}.
In section \ref{sec:SGOPT} we introduce replicas, and we present some of the earliest analytical results obtained applying the replica and cavity methods in the case of two prototypical examples of combinatorial optimization problems, namely the matching and the traveling salesman problems. In section \ref{sec:SAT} we focus on the seminal works \cite{mezard2002analytic,mezard2002random} on the random satisfiability problem. We present the analytical predictions on the phase diagram, as well as the survey propagation algorithm for the solution of given instances of satisfiability. In section \ref{sec:NNs} we move to neural network models, discussing some recent results that build upon a large deviation scheme.

\section{Spin-Glasses and Combinatorial Optimization}
\label{sec:SGOPT}
The earliest attempt to devise a model of SGs can be traced back to the work of Edwards and Anderson \cite{edwards1975theory}. In its simplest form, the Edwards-Anderson model (EA) consists of a random-sign interaction version of the Ising model on a $d$-dimensional regular lattice. As is common practice among statistical physicists, the initial efforts focused on studying a mean-field version of EA, the SK model mentioned before, that is obtained by replacing the $d$-dimensional lattice with a fully connected graph. The goal was to solve the equilibrium thermodynamics of SK, namely to compute the free energy density:
\begin{equation}
\label{eq:freeEn}
f_J=-\lim_{N\rightarrow \infty}\frac{1}{\beta N}\log{Z_{J}},
\end{equation}
where $\beta=1/T$ is the inverse temperature, $Z_J$ is the partition function:
\begin{equation}
Z_J=\sum_{\boldsymbol{s}}e^{-\beta E_J(\boldsymbol{s})},
\end{equation}
and $E_J(\boldsymbol{s})$ is the energy:
\begin{equation}
\label{eq:spinglassdef}
E_J(\boldsymbol{s})=-\sum_{i<j}J_{ij}s_is_j,\quad \boldsymbol{s}=(s_1,\dots,s_N)\in\{\pm 1\}^N.
\end{equation}
The couplings $J_{ij}$'s in \eqref{eq:spinglassdef} are i.i.d.\ Gaussian variables $J_{ij}\sim \mathcal{N}(0,1/N)$. A realization of the $J_{ij}$'s defines a sample of material, in which the couplings are held fixed (quenched disorder), and the only degrees of freedom are the spins $s_i$'s.

It is important to note that for finite \( N \), equation \eqref{eq:freeEn} depends on the specific realization of the couplings. However, in the thermodynamic limit, it converges around its mean value (see \cite{van1982thermodynamic,van1983thermodynamic,petritis1996equilibrium}), denoted as $f=\mathds{E}_{J}f_J$. Consequently, this mean value becomes the focus of analytical computations.

The random sign of the interactions implies frustration: unlike in a standard ferromagnet, for large $N$, it is typically impossible to find a configuration of spins that minimizes all the energetic bonds. In other words, it is impossible to find a configuration where
$J_{ij}s_is_j>0$, for all $i,j$.

The replica method (RM) \cite{brout1959statistical,edwards1975theory,kirkpatrick1978infinite} was the first technique introduced to compute the disorder-averaged free energy $f=\mathds{E}_{J}f_J$. The starting point is the identity: 
\begin{equation}
\mathds{E}_J\log{Z_J}=\lim_{n\rightarrow 0}\frac{1}{n}\log{\mathds{E}_J Z_J^n}.
\end{equation}

The next step is to treat $n$ as an integer. In this case, $Z_J^n$ can be interpreted as the partition function of a total system made of  $n$ independent replicas of the original system. The quantity $\mathds{E}_J Z_J^n$ can then be computed through Gaussian integrations. By averaging over the $J_{ij}$'s, the replica indices become coupled. Because the problem is defined on a fully connected graph, the spatial indices can be factorized, upon introducing an $n\times n$ matrix $\boldsymbol{Q}$  that represents the ``overlaps'' between replicas:
\begin{equation}
\boldsymbol{Q}_{ab}=\frac{1}{N}\sum_{i=1}^Ns_i^{a}s_i^{b},
\end{equation}
where $a,b=1,\dots n$ are replica indices. This results in an expression for the free energy density in terms of an effective action $S_n(\boldsymbol{Q})$ of the form:
\begin{equation}
\label{eq:effActRepl}
f=\lim_{N\rightarrow \infty}-\frac{1}{\beta N}\lim_{n\rightarrow 0}\frac{1}{n}\log{\int d\boldsymbol{Q}\, e^{-N S_n(\boldsymbol{Q})}}.
\end{equation}

The evaluation of \eqref{eq:effActRepl} is a complex problem. A critical aspect is that the effective action $S_n$ is initially computed for integer $n$. Subsequently, an appropriate parametrization of $\boldsymbol{Q}$ is needed, followed by an analytic continuation to real values of $n$ to take the limit $n\rightarrow 0$. After parametrizing $\boldsymbol{Q}$, a feasible and powerful approach is to compute \eqref{eq:effActRepl} using a saddle point approximation on $N$. However, this process necessitates exchanging the limits $n\rightarrow 0$ and $N\rightarrow \infty$.

Since $S_n$ is symmetric under permutations of the replica indices, a natural parametrization for $\boldsymbol{Q}$ is the so-called replica symmetric one, which consists in taking $\boldsymbol{Q}_{ab}=q$ for $a \neq b$. Surprisingly, it turns out that for low temperatures, the RS solution is inconsistent, and the replica symmetry is spontaneously broken (RSB). In \cite{blandin1978theories} and \cite{bray1978replica}, the first attempts to find the correct parametrization of $\boldsymbol{Q}$ in the RSB phase were presented. However, the correct ansatz was found by Parisi shortly after (see \cite{parisi1979infinite,parisi1979toward,parisi1980magnetic,parisi1980sequence,parisi1980order}). Parisi showed that the RSB phase requires the introduction of an infinite number of order parameters, in the form of a function $q(x)$ on the interval $[0,1]$. This function is determined by a variational principle for the free energy (the Parisi formula), whose functional is evaluated by solving a partial differential equation for each trial $q(x)$. Some years later, a physical interpretation of the Parisi order parameter was presented in \cite{parisi1983order}, building upon the many-valleys picture \cite{kirkpatrick1978infinite,de1980white,bray1981metastable1,sompolinsky1981time,young1981tap,palmer1982broken}. In the RSB phase, ergodicity is broken, and the partition function decomposes into a sum of contributions from the pure states:
\begin{equation}
Z = \sum_{\alpha} e^{-\beta N f_{\alpha}},
\end{equation}
where $f_{\alpha}$ is the free energy density of pure state $\alpha$. It is shown that the $q(x)$ introduced by Parisi allows for the computation of the probability distribution of the overlaps between these pure states. A striking geometrical property of the space of pure states is their ultrametric topology: picking three states with mutual overlaps $q_1\geq q_2\geq q_3$, one finds $q_1\geq q_2=q_3$ with high probability \cite{mezard1984nature}.

The predictions of the RM were confirmed rigorously several years later \cite{guerra2002thermodynamic,talagrand2006parisi,panchenko2013sherrington}.

In \cite{mezard1986skmodel} an alternative technique to replicas, the \emph{cavity} method (CM) which generalizes some known mean field techniques to the case of disordered systems, is proposed to solve the SK model. Interestingly, the CM is capable of reproducing all the results of the replica symmetry-breaking solution of SK, with the advantage of bypassing the technical complications deriving from the limit $n \rightarrow 0$. The basic idea is to consider an auxiliary graph, obtained by local modifications of the original one, such as the addition of a ``cavity'' spin. It is shown that, due to the mean-field nature of the problem, it is possible to model the influence that the rest of the system has on the cavity spin in a self-consistent way. Remarkably, in contrast to the RM, the self-consistent equations deriving from the CM are closed on physical quantities.

In the 80s it became evident that the replica and the cavity methods could be also used as analytical tools to study average properties of combinatorial optimization problems (COP)s, after realizing that COPs can be formulated as problems in statistical mechanics. This approach turned out to be very fruitful, due to the generality of the geometrical picture evoked by the configuration space structure of SK. 
It is possibly fair to say that these methods were (and partially still are) the only known ones that lead to quantitative predictions for highly non-trivial non-convex optimization problems. New mathematical tools and new algorithms have emerged from these methods.

We define a class of combinatorial optimization problems to be a set of instances $\mathcal{I}$. A given instance $I \in \mathcal{I}$ specifies: i) a space of feasible solutions $F(I)$, where each element $C \in F(I)$ is a bit string; ii) a cost function $E_I$, which maps $F(I)$ into $\mathds{R}$. Given $I \in \mathcal{I}$, the goal is to find an assignment $C \in F(I)$ that minimizes $E_I(C)$.

Computational complexity theory categorizes COPs based on the worst-case resources required to solve them, such as time and space. In this classification, \textsf{NP} (Nondeterministic Polynomial time) problems can be informally defined as those for which a potential solution can be efficiently checked for correctness in the worst case, where ``efficiently'' means requiring a time that is at most polynomial in the size of the instance. A fundamental role is played by \textsf{NP}-complete problems. The existence of an algorithm for solving efficiently in the worst case an \textsf{NP}-complete problem would imply the existence of efficient algorithms for all elements of \textsf{NP}. However, it is conjectured that no such algorithm exists, namely that \textsf{NP}$\neq$\textsf{P}, the subclass of \textsf{NP} for which polynomial time algorithms exist. The best known algorithms are able to solve \textsf{NP}-complete problems with a time cost growing exponentially in the size of the system in the worst cases. In \cite{jerrum1993ising}, it is shown that no deterministic polynomial-time algorithm can approximate the partition function of a general Ising spin glass within a factor of two in the worst case, unless $\mathrm{P}=\mathrm{NP}$. 

The fundamental observation making the link with statistical physics is to read the instances as samples of a physical system, the space of feasible solutions as its configuration space, and the cost function as an energy. This connection was exploited for the first time numerically with the introduction of ``Simulated Annealing'', in the seminal work \cite{kirkpatrick1983optimization}. The idea is to define a fictitious temperature $T=1/\beta$, to assign a Boltzmann weight $e^{-\beta E(C)}$ to each configuration $C$ of the variables, and to perform Monte Carlo steps in combination with a suitable annealing protocol on the temperature $T$, in order to sample low-energy configurations. Simulations performed with simulated annealing on the famed Traveling Salesman Problem (TSP) gave evidence for properties reminiscent of SGs, namely a hierarchical and ultrametric structure of the configuration space \cite{kirkpatrick1985configuration}.

\subsubsection*{Matching and Traveling Salesman Problems}
\begin{figure}
    \centering
    \includegraphics[width=0.8\textwidth]{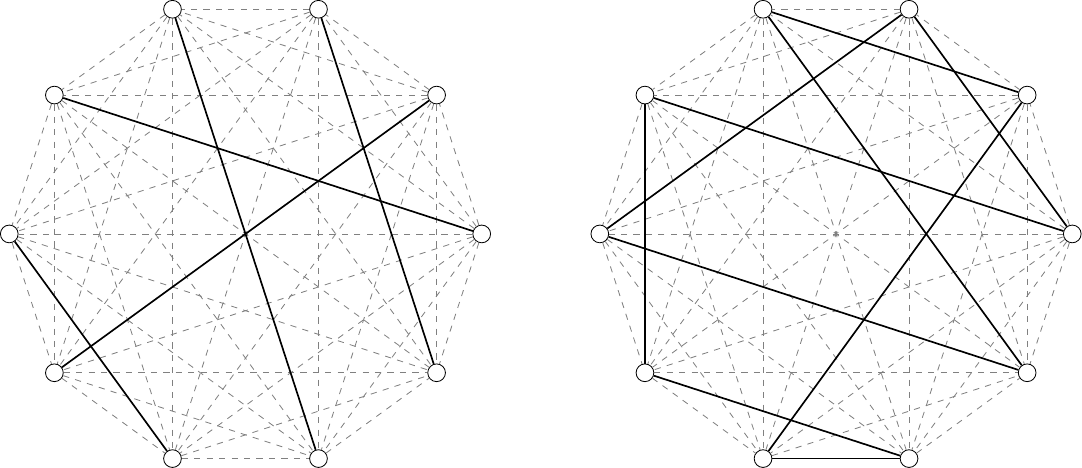}
    \caption{Representation (continuous lines) of a perfect matching on the left, and a Hamiltonian cycle (which is a feasible solution in a TSP instance) on the right.}
    \label{fig:MPTSP}
\end{figure}
The matching problem (MP) and the Traveling salesman problem (TSP) (see Fig.~\ref{fig:MPTSP}) are the first COPs studied analytically with statistical physics techniques. Let us start with the simplest problem, matching. In the seminal paper \cite{M_zard_1985} a replica study of MP on the fully-connected graph $K_{2N}$ is presented. In MP one is given $2N$ points, and a matrix of ``distances'' $l_{ij}\geq 0$, $i,j=1,\dots,2N$ associated with the edges of $K_{2N}$, and is asked to find a perfect matching with minimum length, i.e.\ a set of $N$ edges such that: i) every couple of edges belonging to the matching is node-disjoint; ii) the sum of the lengths of the edges of the matching is minimum. 
Note that referring to the $l_{ij}$'s as distances might be misleading in this case, indeed the disorder is uncorrelated, implying that the triangle inequality is generally not satisfied. For each pair $i,j$, define $n_{ij}=n_{ji}\in\{0,1\}$, with $n_{ii}\equiv 0$. The occupation $n_{ij}$ is one if the pair belongs to the matching, and zero otherwise, and the matching condition reads:
\begin{equation}
\sum_{j=1}^{2N}n_{ij}=1,\quad \forall i. 
\end{equation}
The cost (the energy) one wants to minimize is
\begin{equation}
E(\{n_{ij}\})=\sum_{i<j}l_{ij}n_{ij}.
\end{equation}
An instance of MP is specified by the matrix of distances $l_{ij}$. In the statistical physics language this matrix represents the unique source of quenched disorder. An obvious question, given an instance of MP, is to find a minimum weight perfect matching. For MP, there exists an efficient algorithm (the Blossom algorithm by Edmonds) that, in the worst case, solves the optimization problem in polynomial time \cite{edmonds1965paths}. The replica study in contrast is aimed at studying the instance averaged properties of the optimal solutions, e.g.\ the computation of the average optimal cost density $\hat{\mathcal{E}}$, properly rescaled with $N$, of the minimum weight perfect matching on an ensemble of instances. In \cite{M_zard_1985} the distances $l_{ij}$ are taken as i.i.d.\ random variables extracted from a probability distribution $\rho(w)$ of the form: 
\begin{equation}
\label{eq:FormOfDistr}
\rho(w)\stackrel{w\rightarrow 0}{\approx}w^r,\quad r>-1.
\end{equation}
Note that the length-scale of a link belonging to the matching is fixed by the typical distance of the ``nearest'' neighbor, and
being the topology fully connected, in the large $N$ limit this distance tends to zero. For this reason the only relevant property of the weights distribution $\rho(w)$ in the thermodynamic limit is its behavior around $w=0$. For the computation of $\hat{\mathcal{E}}$, the authors define a partition function at inverse temperature $\beta$:
\begin{equation}
Z_l(\beta)=\sum_{\{n_{ij}=0,1\}}\prod_{i=1}^{2N}\delta\left(1-\sum_{j}n_{ij}\right)e^{-\beta E(\{n_{ij}\})},
\end{equation}
and study with the replica method the zero temperature limit of the disorder-averaged free energy. However, to achieve the right limit, it is necessary to rescale the temperature in such a way that the energy is an extensive quantity \cite{vannimenus1984statistical}. As each vertex has $\mathcal{O}(N)$ neighbors, one may expect that the typical weight $w^{\star}$ belonging to the matching satisfies:
\begin{equation}
\label{eq:scalaComeunoSuN}
\int_{0}^{w^{\star}}dw\rho(w)=\mathcal{O}\left(\frac{1}{N}\right),
\end{equation}
implying that $w^{\star}=\mathcal{O}(N^{-\frac{1}{r+1}})$. Therefore, the energy scales like $N^{1-\frac{1}{r+1}}$, and to obtain an extensive energy in the large $N$ limit, one has to rescale $\beta$ as follows:
\begin{equation}
\beta=\hat{\beta}N^{\frac{1}{r+1}}.
\end{equation}
Thus, in order to compute $\hat{\mathcal{E}}$, one has to study the thermodynamic limit at fixed $\hat{\beta}$, and then take the limit $\hat{\beta}\rightarrow\infty$ to study the properties of the ground state:
\begin{equation}
\hat{\mathcal{E}}=\lim_{\hat{\beta}\to\infty}\lim_{N\rightarrow \infty}-\frac{1}{\hat{\beta}N}\mathds{E}_l\log{Z_l\left(\hat{\beta}N^{\frac{1}{r+1}}\right)}.
\end{equation}
In \cite{M_zard_1985,mezard1987solution}, M\'ezard and Parisi use the replica method under the replica symmetric assumption to follow this program in the case of the MP defined before, and for another variant, the assignment problem, in which the graph is complete and bipartite. In particular, for $r=0$ they find
\begin{equation}
\label{eq:AsymptoticCost}
\hat{\mathcal E}_{\rm MP}=\frac{\pi^2}{12},\qquad \hat{\mathcal E}_{\rm AP}=\frac{\pi^2}{6},
\end{equation}
for the non-bipartite matching problem and for the bipartite assignment problem, respectively.

Interestingly, the RS solution of MP is more complicated than the SK one. At the RS level, the order parameter in SK is fixed by a single number, namely the width of the Gaussian describing the probability distribution of the local field acting on a spin, being the local field given by the sum of $N$ independent random variables. In MP, despite the graph being still fully connected, each variable interacts with a small number of its neighbors, since the typical length of the edges belonging to the matching scales like $N^{-\frac{1}{r+1}}$. Therefore the effective graph of interaction results diluted, and the distribution of the local field is non-trivial. 

The validity of replica predictions on MP, was corroborated by numerical simulations, and the consistency of the RS ansatz has been checked in \cite{mezard1987solution,Parisi_2002} by a study of the stability of the saddle point. In particular, for distributions with $r=0$, it is shown that the spectrum of the Hessian matrix contains only non-negative eigenvalues at any temperature. 

In 2001 Aldous \cite{Aldous_2001} provided a rigorous treatment of the random link matching problem on complete bipartite graph, in the limit $N\rightarrow\infty$, and for independently and exponentially distributed random weights, confirming the prediction of Eq.~\eqref{eq:AsymptoticCost}. 

In \cite{mezard1987solution,Parisi_2002} the replica analysis of MP is also extended to the computation of finite-size corrections of the average energy density. The sub-leading corrections, unlike the thermodynamic average energy density, depend on the expansion of the probability distribution of the lengths up to, at least, the second term \cite{Parisi_2002}.

In \cite{Parisi_2002} for the complete non bipartite graph, in the case of weights distributed according to the uniform law on the interval $[0,1]$, Parisi and Ratiéville found:
\begin{equation}
\label{eq:flatRLMP}
\hat{\mathcal{E}}^N_{\textnormal{flat}}=\frac{\pi^2}{12}+\frac{1}{N}\left(-\frac{\zeta(3)}{2}+\sum_{p=0}^{\infty}\frac{I_{2p+1}}{2p+1}\right)+o\left(\frac{1}{N}\right),
\end{equation}
where $I_p$ are integrals in $p$ variables that can be computed analytically. In the same work, by using the flat distribution in the bipartite case, they also found:
\begin{equation}
\hat{\mathcal{E}}^{N}_{\textnormal{flat}}=\frac{\pi^2}{6}-\frac{1}{N}(1+2\zeta(3))+o\left(\frac{1}{N}\right).
\end{equation}

In 1998 Parisi conjectured \cite{parisi1998conjecture} an exact formula for the averaged optimal cost of the assignment problem on $K_{N,N}$ as a function of the size $N$ of the system, $\hat{\mathcal{E}}^{N}$, in the case of i.i.d.\ lengths exponentially distributed:
\begin{equation}
\label{eq:exponentialDistribution}
\rho(w)=e^{-w}\quad w>0.
\end{equation}
The conjecture is the following:
\begin{equation}
\label{eq:costAssignmFunctionOfN}
\hat{\mathcal{E}}^{N}=\sum_{j=1}^{N}\frac{1}{j^2}.
\end{equation}
By taking the limit $N\rightarrow\infty$ in \eqref{eq:costAssignmFunctionOfN}, the result of \eqref{eq:AsymptoticCost} is recovered:
\begin{equation}
\lim_{N\rightarrow\infty}\hat{\mathcal{E}}^{N}=\frac{\pi^2}{6}.
\end{equation}
The Parisi's conjecture given in \eqref{eq:costAssignmFunctionOfN} has been
independently proved by Linusson and W\"astlund \cite{Linusson_2004} and Nair, Prabhakar, and Sharma \cite{Nair_2005} in 2004. 

The second problem studied by the replica and the cavity method was the TSP, a much harder problem than MP. At variance with MP, the TSP is  NP-complete \cite{garey1979computers}, and therefore there are no known polynomial algorithms that can solve it optimally in the worst case. As already mentioned, the study of TSP from a statistical physics perspective initiated with the numerical works on Simulated Annealing \cite{kirkpatrick1983optimization} \cite{kirkpatrick1985configuration}. The first theoretical analyses concerning the average properties on the fully connected graph $K_N$ with replicas were presented in \cite{orland1985mean}, \cite{mezard1986replica} and \cite{mezard1986mean}. Given $N$ points associated with the vertices of $K_N$, and a set of lengths $l_{ij}$ defined on its edges, TSP asks for a Hamiltonian tour of minimal length, i.e. a connected closed path passing through all nodes just once, in which the sum $L_l(\text{tour})$ of the $l_{ij}$'s belonging to the path is minimal. Taking the $l_{ij}$'s i.i.d. according to \eqref{eq:FormOfDistr}, and following the same reasoning of MP for the rescaling of $\beta$ with $N$, one can write the partition function:
\begin{equation}
\label{eq:PartFunctTSP}
Z_l(\hat{\beta})=\sum_{\text{tours}}e^{-\hat{\beta}N^{\frac{1}{r+1}}L_l(\text{tour})}. 
\end{equation}
At variance with MP, due to its non-locality, the Hamiltonian constraint on the path cannot be simply enforced introducing binary occupation variables on the edges. In \cite{orland1985mean} it is shown that \eqref{eq:PartFunctTSP} can be written in terms of local interactions at the price of defining an auxiliary partition function involving $m$-components vectors on the nodes of the graph. For the same reasons of MP, also the TSP order parameter is a non-trivial function. Indeed also in this case each node effectively interacts with a small number of its neighbors. An approximate solution to the saddle-point equations for the order parameter in the RS ansatz were found in the case of flat distribution ($r=0$) in \cite{mezard1986replica}. 

In \cite{mezard1986mean} and \cite{krauth1989cavity} the random TSP is studied within the framework of the cavity method. In particular in \cite{krauth1989cavity}, Krauth and Mézard manage to solve the cavity equations directly at zero temperature. Interestingly, random TSP turns to be an example where, despite the equivalence of both methods, the cavity equations are much easier to solve than the replica ones. This is a consequence of the fact that the cavity method deals directly with physical quantities, like the distribution of the local fields acting on a variable \cite{mezard1989optimization}. The result for the average minimum length $\hat{L}$ of the tour for $r=0$ and $r=1$, is respectively:
\begin{equation}
\label{eq:reResults}
\hat{L}_{r=0}=2.0415\dots,\quad \hat{L}_{r=1}=1.8175\dots\sqrt{N}.
\end{equation}
In the same works, \eqref{eq:reResults} are shown to be in agreement with numerical estimates obtained with efficient heuristic algorithms.

\section{From Replicas to Cavity to Algorithms}
\label{sec:SAT}
The 1990s marked a paradigm shift, as the focus started to move from studying ensemble averages to designing efficient algorithms that leverage the theoretical understanding of the RSB phenomenology for solving single instances of COPs. This new quest was motivated by the discovery that algorithms used to solve random instances of \textsf{NP}-complete problems display a critical behavior associated with a huge variability of running times. In particular, it was observed that for large systems, random instances are hard to solve only near specific ``critical'' values of the control parameters defining the disorder ensemble
\cite{kirkpatrick1994critical,hogg1996phase,monasson1999determining,dubois2001special}. The average-case complexity perspective gained from the explorations of random instances of COPs presented new challenges of significant interest to physicists and computer scientists. These include understanding the onset of exponential regimes in algorithms, developing new mathematical tools to handle random instances, and exploring applications in coding theory and post-quantum cryptography. 

The RSB theory offers a framework to address some of these fundamental questions. In this next section, we discuss the seminal results achieved within the context of this theory through the study of a paradigmatic \textsf{NP}-complete problem: the Boolean satisfiability problem.

\subsection{Random KSAT}
The satisfiability problem asks whether it is possible to simultaneously satisfy a set of $M$ constraints among N Boolean variables, where each constraint is a clause formed by the logical OR of a subset of variables or their negations. An instance is called satisfiable (SAT) if it is possible to satisfy all constraints, and unsatisfiable (UNSAT) otherwise. Let's represent each variable with a spin $s_i$, $i=1,\dots,N$, in such a way that $s_i=1$ if the variable is \texttt{TRUE} and $s_i=-1$ if it is \texttt{FALSE}. We denote by $\boldsymbol{s}=(s_1,s_2,\dots,s_N)$ a configuration of the system. Let's also introduce for each variable $i$ entering clause $a$, $a=1,\dots,M$, denoted with $i\in\partial a$, a ``coupling'' $J^a_i=1$ if $i$ appears negated in $a$, or $J^a_i=-1$ otherwise. With this notation the goal is to find, or to prove that it does not exist, a configuration $\boldsymbol{s}$ of the spins with zero energy $E(\boldsymbol{s})=0$: 
\begin{equation}
E(\boldsymbol{s})=\sum_{a=1}^M E_a(\boldsymbol{s}_{\partial a};\boldsymbol{J}^a),\quad E_a(\boldsymbol{s}_{\partial a};\boldsymbol{J}^a)=2\prod_{i\in\partial a}\frac{1+J^{a}_is_i}{2},
\end{equation}
where $\boldsymbol{s}_{\partial a}=\{s_i,i\in\partial a\}$, and $\boldsymbol{J}^a=\{J_i^a,i\in\partial a\}$. Note that $E(\boldsymbol{s})$ simply counts twice the number of constraints (clauses) that are violated. Let's restrict to the case in which each clause contains a fixed number $K$ of variables (KSAT). For $K\geq 3$, KSAT is \textsf{NP}-complete \cite{cook1971complexity},\cite{levin1973universal}.
In random KSAT one extracts uniformly, independently and with replacement $M=\alpha N$ clauses of $K$ variables in the set of all $2^K\binom{N}{K}$ possible ones. A formula generated in this fashion can be represented graphically with a factor graph, namely a bipartite graph in which ``square''-nodes are associated with clauses, and ``circle''-nodes with variables. An edge between circle $i$ and square $a$ is added if and only if variable $i$ enters clause $a$. Interestingly the ensemble of random factor graphs defined by the formula generation process discussed above is of the Erd\H{o}s--R\'enyi type. 

Friedgut \cite{friedgut1999sharp} proved the existence of a threshold sequence $\alpha_c(N)$ such that, for every fixed $\varepsilon>0$, the probability that a random instance is SAT tends to one at clause density $\alpha_c(N)-\varepsilon$ and to zero at clause density $\alpha_c(N)+\varepsilon$, as $N\to\infty$. The transition is therefore asymptotically sharp, although the theorem does not establish convergence of $\alpha_c(N)$ to a limiting value.

From a topological perspective, in the limit of large instances with $\alpha=O(1)$, random formulas acquire peculiar properties. Indeed, the distribution of the number of clauses connected to a given variable becomes a Poissonian with mean $\alpha K$, implying that each variable has a finite number of neighbors and a notion of distance is defined in the model. Moreover, the neighborhood of a variable taken at random is typically a tree up to a distance that is diverging with $N$ (as $\log N$). Such ``locally-tree-like'' random graphs are known as Bethe lattices (BL) in the physics community. These models are evidently different from the fully connected ones, and are more challenging to solve analytically due to their spatial structure. Efforts to extend the replica method to the Bethe lattice setting resulted, when introducing RSB effects, in complicated saddle-point equations that could be handled only by variational approximations \cite{dominicis1987replica,wong1988intensively,monasson1996entropy,monasson1998optimization,biroli2000variational}. A crucial contribution is found in \cite{M_zard_2001,mezard2003cavity}, where Mézard and Parisi show how to generalize the cavity method on BLs using the simplest ansatz for breaking the replica symmetry, known as the one-step replica symmetry breaking (1RSB) solution. Interestingly it is found that the cavity equations, at variance with the replica ones, could be solved non-perturbatively with simple numerical techniques. 

Building upon this new RSB cavity method, in the seminal papers \cite{mezard2002analytic,mezard2002random}, random KSAT is studied, addressing two types of fundamental questions. The first is theoretical, focusing on the computation of typical properties of random instances, such as determining $\alpha_c$ and understanding why instances near $\alpha_c$ in the SAT phase are difficult to solve. The second involves developing an efficient algorithm capable of solving specific instances. For the first time it was shown that the RSB cavity method could be used for this purpose, and not only for the computation of average thermodynamic quantities. In the following we want to review some of these important results.

Let's start with the definition of pure states. At zero temperature in KSAT, a pure state can be defined as a cluster of isoenergetic configurations connected by single spin flips, that is ``locally stable'', meaning its energy cannot be decreased by flipping any finite number of spins \cite{mezard2002random}. A qualitative picture of the phase diagram found in \cite{mezard2002analytic,mezard2002random} for $K=3$ is the following. For low enough $\alpha$, the thermodynamics is described by a unique pure state at zero energy, and the problem is replica symmetric. For $\alpha_d<\alpha<\alpha_c$, where $\alpha_d$ is called \emph{dynamic} transition, the phase space breaks into an exponential number of pure states, and the description of the thermodynamics requires RSB. In this region, typical instances are still SAT. Above $\alpha_c$, the SAT-UNSAT transition, typical formulas are UNSAT. This picture results from the so-called \emph{energetic} cavity method \cite{mezard2003cavity}. This is the cavity method formulated at zero temperature, assuming that the entropic contribution to the free energy is negligible for $T\rightarrow 0$. Let us define two \emph{conditioned} ``cavity'' energies: $E_{i \rightarrow a}(s_i)$ and $\hat{E}_{b \rightarrow i}(s_i)$. The former is obtained by removing clause $a$, and optimizing the energy over all spins except $i$, conditioned on $s_i$. The latter is obtained by removing all clauses in which $i$ appears, except clause $b$, and again optimizing the energy over all spins except $i$, conditioned on $s_i$. Due to the locally-tree-like property of the factor graph, when removing a clause $a$ at random, the variables in $\partial a$ become typically far apart. Therefore if the thermodynamic state is unique one can consider the cavity spins as independent  and write the following RS cavity equations:  
\begin{equation}
\label{eq:RScavEq1}
E_{i\rightarrow a}(s_i)=\sum_{b\in\partial i\setminus a}\hat{E}_{b\rightarrow i}(s_i),
\end{equation}
\begin{equation}
\label{eq:RScavEq2}
\hat{E}_{a\rightarrow i}(s_i)=\min_{\underline{s}_{\partial a\setminus i}}\left[E_a(\underline{s}_{\partial a})+\sum_{j\in\partial a\setminus i}E_{j\rightarrow a}(s_j)\right].
\end{equation}
Since the spins are binary variables  the cavity energies can be simply parametrized in terms of cavity fields $h_{i\rightarrow a}$ and cavity biases $u_{b\rightarrow i}$:
\begin{equation}
\label{eq:paramEn}
E_{i\rightarrow a}(s_i)\doteq -h_{i\rightarrow a}s_i,\quad \hat{E}_{b\rightarrow i}(s_i)\doteq -u_{b\rightarrow i}s_i,
\end{equation}
where ``$\doteq$'' is the equality up to an additive constant. Using definitions \eqref{eq:paramEn} in \eqref{eq:RScavEq1} and \eqref{eq:RScavEq2}, one finds self-consistent equations for the cavity fields and cavity biases that can be closed on $u_{a\rightarrow i}\in\{0,-J^{a}_i\}$ and $h_{i\rightarrow a}\in\mathbb{Z}$, obtaining the so-called ``warning propagation'' equations:
\begin{equation}
\label{eq:cavEqsfields}
h_{i\rightarrow a}=\sum_{b\in\partial i\setminus a}u_{b\rightarrow i},\quad u_{a\rightarrow i}=\hat{u}\big(\{h_{j\rightarrow a}\}_{j\in\partial a\setminus i};\boldsymbol{J}^a\big)\equiv -J^a_i\prod_{j\in\partial a\setminus i}\theta\left(J^{a}_jh_{j\rightarrow a}\right).
\end{equation}
Here, $\theta(x)=1$ for $x>0$, and $\theta(x)=0$ otherwise.
The cavity biases $u_{a\rightarrow i}$, also called ``warnings'', have a simple meaning. If $u_{a\rightarrow i}=-J^a_i$, then spin $i$ is ``warned'' that clause $a$ is satisfied only if $s_i=-J^a_i$. If $u_{a\rightarrow i}=0$, $s_i$ does not receive any warning from $a$. In the RS phase Eqs.~\eqref{eq:cavEqsfields} admit a unique solution, the sets of fixed-point ``messages'' $\{h_{i\rightarrow a}\}$, $\{u_{a\rightarrow i}\}$. The cavity estimate for the assignment of a variable $i$ is then given by the magnetization $m_i=\text{sign}\left(\sum_{a\in\partial i}u_{a\rightarrow i}\right)$. 

If the replica symmetry is broken, the RS cavity equations admit many fixed points, each of them associated with a pure state. The 1RSB cavity method assumes that in the thermodynamic limit the number of states at a given energy density $e=E/N$ behaves like $e^{N\Sigma(e)}$. One then introduces a zero-temperature 1RSB ``free energy'' $f_{1rsb}(y)$ given by:
\begin{equation}
\label{eq:1rsbF}
e^{-y N f_{1rsb}(y)}=\int de\,e^{N(\Sigma(e)-ye)},
\end{equation}
where the parameter $y$ plays the same role of an inverse temperature. In particular by choosing larger values of $y$, one selects states of lower energy, until a maximum value $y^*$ corresponding to the global ground states. In the SAT phase the global ground states, that in this region are the actual solutions of KSAT formulas, are obtained as a ``zero-temperature'' limit, namely $y^*=\infty$. In order to compute \eqref{eq:1rsbF}, one needs to characterize the statistical properties of the cavity fields and cavity biases, that in the RSB phase depend on the state. Given the directed edge $a\rightarrow i$, let us introduce the probability distribution (survey) of the cavity bias in the different states:
\begin{equation}
Q^{(y)}_{a\rightarrow i}(u)\propto \sum_{\omega}\delta(u-u^{\omega}_{a\rightarrow i}),  
\end{equation}
where the sum over $\omega$ is taken over the states having the energy density $e$ selected by the parameter $y$. In \cite{mezard2003cavity} it is shown that subject to an assumption on the distribution of the energies of the pure states, it is possible to write self-consistent equations for the surveys at the 1RSB level. The (survey propagation) self-consistent equations take the form:
\begin{equation}
\label{eq:SPeq}
Q^{(y)}_{a\rightarrow i}=\mathcal{J}_y\left(\big\{Q^{(y)}_{b\rightarrow j},\,b\neq a\,\,\text{and}\,\, j\in\partial a\setminus i\big\}\right),
\end{equation}
where $\mathcal{J}_y$ is a non-linear integral operator that depends on $y$. Note that, since $u_{a\rightarrow i}\in\{0,-J^a_i\}$, each survey $Q^{(y)}_{a\rightarrow i}$ can be simply parametrized in terms of a single number $\eta_{a\rightarrow i}$, the probability that on $a\rightarrow i$ no warning is sent on the pure states selected by $y$:
\begin{equation}
Q^{(y)}_{a\rightarrow i}(u)=\eta_{a\rightarrow i}\delta(u)+(1-\eta_{a\rightarrow i})\,\delta(u+J^a_i),
\end{equation}
implying that \eqref{eq:SPeq} becomes an iterative rule for $\eta_{a\rightarrow i}$.

\begin{figure}
    \centering
    \includegraphics[width=0.8\textwidth]{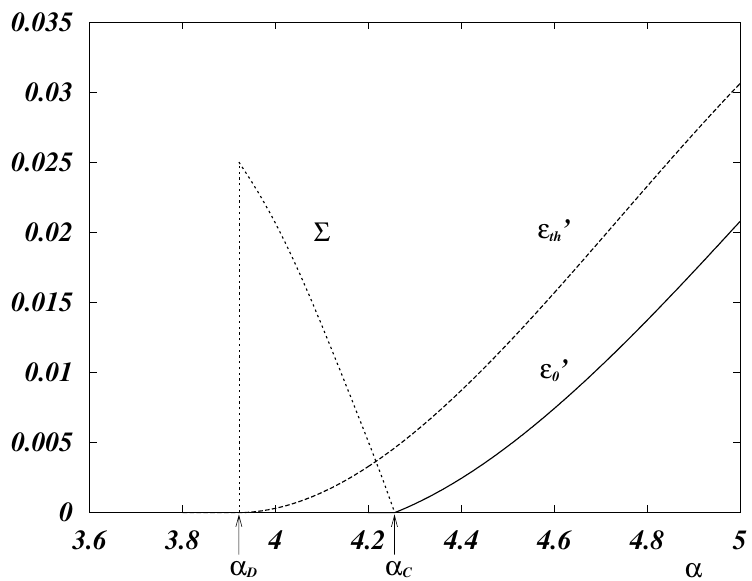}
    \caption{Phase diagram of the random 3SAT problem. $\Sigma$ represents the complexity of the pure states with zero energy (solutions). The continuous line ($\epsilon_0'$) represents the ground states energy density divided by two, and the dashed line ($\epsilon_{th}'$) represents the threshold energy density, i.e.\ the energy of the most numerous pure states, divided by two. The figure is taken from \cite{mezard2002random}.}
    \label{fig:PhaseDiagramKSAT}
\end{figure}

Although Equation \eqref{eq:SPeq} is formulated for a single sample, it is possible to derive a closed equation for the entire order parameter of the problem, represented by the probability distribution $P^{(y)}(\eta)$ of the surveys. The knowledge of $P^{(y)}(\eta)$ allows to compute $f_{1rsb}$ and the complexity $\Sigma$. In \cite{mezard2002analytic,mezard2002random} the equation for $P^{(y)}(\eta)$ is solved numerically by population dynamics algorithms, leading to the following predictions for $\alpha_d$ and $\alpha_c$ (see Fig. \ref{fig:PhaseDiagramKSAT}). The dynamical transition $\alpha_d \approx 3.921$ is identified as the value of $\alpha$ below which the solution is ``paramagnetic'', $P(\eta)=\delta(\eta-1)$. For $\alpha<\alpha_d$, typical instances are SAT, and solutions can be found even by a greedy energy minimization, like a zero temperature Metropolis. For $\alpha_d<\alpha<\alpha_c$, the 1RSB order parameter and the complexity $\Sigma(e)$ become non-trivial. The computation of $\Sigma(e)$ shows that there is an exponential number of pure states with zero energy, therefore typical instances are still SAT. The authors of \cite{mezard2002analytic,mezard2002random} call this region Hard-SAT-Phase (HSP), since despite being typical instances SAT, the presence of an exponential number of states with non-zero energy  makes it difficult to find a solution. Indeed it is found that in Hard-SAT-Phase greedy algorithms get usually trapped into the most numerous states, those with energy density $\epsilon_{th}>0$ maximizing $\Sigma(e)$, and a large class of MCMC sampling algorithms suffers from an exponential slowing-down above $\alpha_d$. In \cite{mezard2002analytic,mezard2002random} the SAT-UNSAT transition is identified as the point where the complexity of solutions becomes zero, predicting $\alpha_c\approx 4.267$ for $K=3$, in agreement with numerical simulations \cite{crawford1996experimental}. In \cite{mertens2006threshold}, expanding the 1RSB cavity equations, it is predicted that for large $K$:
\begin{equation}
\label{eq:SATUNSATasymptotics}
\alpha_c(K) = 2^K \log{2}-\frac{1}{2}(1+\log{2})+o\left(1\right).
\end{equation}
Asymptotic formula \eqref{eq:SATUNSATasymptotics} was proved in \cite{ding2015proof}.

One of the most important contributions of \cite{mezard2002analytic,mezard2002random} is the design of an algorithm for satisfiability taking advantage of the theoretical framework we just introduced. The solution of the survey propagation equation on a given sample of the problem allows to compute for each site $i$ the probability distribution $P_i(H)$ of the (integer) local field $H_i=\sum_{a\in\partial i}u_{a\rightarrow i}$ (remember that $H_i$ depends on the state), having the following simple interpretation. The cumulatives $w_i^{+}=\sum_{H=1}^{\infty}P_i(H)$, and $w_i^{-}=\sum_{H=-\infty}^{-1}P_i(H)$, represent the fraction of the number of states (having energy selected by the chosen value of $y$) in which, respectively, $s_i=1$ and $s_i=-1$. In \cite{mezard2002analytic,mezard2002random} this important information is used within a ``decimation'' scheme that the authors name survey inspired decimation (SID). Given an instance of the problem, the survey propagation equation are solved by iteration for a sufficiently large value of $y$, and for each site $i$ the probabilities $w^{+}_i,w^{-}_i$ are computed. Then a specified fraction of the most polarized variables, namely those with largest $|w_i^{+}-w_i^{-}|$ are fixed to $\text{sign}(w_i^{+}-w_i^{-})$. At this stage, the instance is simplified: all satisfied clauses are erased, and all fixed variables are removed from the graph. There may still be ``unit'' clauses, i.e., clauses with a single variable. These clauses are automatically satisfied by assigning the correct value to the variable, and the formula is then simplified again. At this point, the procedure is repeated on the simplified formula. Remarkably, SID efficiently found satisfying assignments for all computer science benchmarks of hard 3SAT formulas published at the time, as demonstrated in \cite{mezard2002analytic}, with instances having $N = 2000$. In the same paper, the authors successfully tested SID on much larger 3SAT instances with $N=10^5$ in the Hard-SAT-phase (up to $\alpha\approx 4.2$). 

These seminal works on KSAT marked a significant advancement in the interdisciplinary collaboration between physics and computer science \cite{gomes2002satisfied} \cite{mezard2003passing} \cite{knuth2015art}, and contributed to pave the way for research on the geometry of solution spaces in hard random COPs. Years later, a detailed picture of the ``structural'' phase transitions in the space of solutions of random COPs, considering the internal entropy of clusters, was presented in \cite{krzakala2007gibbs} (see also \cite{mezard2005clustering,mezard2005landscape,achlioptas2006solution,montanari2008clusters}). This more refined analysis unveils the existence of other phase transitions. For $K\geq 4$, a ``condensation'' threshold $\alpha_{cond}$ is identified, with $\alpha_d<\alpha_{cond}<\alpha_c$, such that for $\alpha_{cond}<\alpha<\alpha_c$ the uniform measure over solutions becomes dominated by a sub-exponential number of clusters. Much of the aforementioned understanding of the solution space has been rigorously validated in the large-$K$ limit \cite{achlioptas2008algorithmic,achlioptas2006solution}. 

In \cite{parisi2003backtracking,marino2016backtracking}, it is shown that SID can be enhanced with a ``backtracking'' strategy, which involves appropriately changing the assignment of previously fixed variables during the iteration. 

Interestingly, what the authors of \cite{mezard2002analytic,mezard2002random} originally termed the Hard-SAT phase, is indeed a hard phase for algorithms sampling solutions uniformly. However algorithms like SID can efficiently find solutions in a window above $\alpha_d$. At the same time for large values of $K$, it is shown that the best known algorithms stall at $\alpha_{alg}\approx 2^K/K\log{K}$, that is strictly smaller than $\alpha_c(K)$ (see Eq.~\eqref{eq:SATUNSATasymptotics}). Therefore there is evidence of an algorithmic gap: a region of $\alpha$ in which instances are typically SAT, but even the best known algorithms fail at finding efficiently satisfying assignments. Building upon these pioneering works in statistical physics, the geometric explanation of algorithmic gaps in COPs remains a highly studied and active topic today \cite{gamarnik2022disordered}. It was conjectured (see \cite{mezard2005clustering,achlioptas2006solution,marino2016backtracking}) based on these foundational studies that finding solutions in clusters containing a backbone of ``frozen'' spins, namely a subset of variables that can take a unique assignment inside the cluster \cite{monasson1999determining,semerjian2008freezing,braunstein2016large}, is algorithmically hard. Such ``frozen'' clusters are shown to appear in KSAT in a window comprised between $\alpha_d$ and $\alpha_c$ \cite{marino2016backtracking,montanari2008clusters}. 
Therefore it is expected that a necessary condition for the success of the smartest algorithms is the existence of possibly sub-dominant pure states not containing frozen variables. However this condition is not sufficient \cite{baldassi2015subdominant,gamarnik2021overlap,gamarnik2022algorithms}. This fundamental point is further discussed in Section \ref{sec:NNs}, where we review a hardness criterion based on a large-deviation replica method in the context of artificial neural networks. 

To conclude this section, we mention that message passing algorithms of the survey propagation type find concrete applications outside the realm of physics and computer science. An example is \cite{weigt2009identification}, where a message passing algorithm is used to infer protein contacts from sequences of data.

\section{The geometry of the space of solutions in Neural Networks}
\label{sec:NNs}

In this section, we will explore certain properties of neural devices that shed light on the algorithmic significance of rare regions in the energy landscape. Investigating these properties necessitates a specific large deviation analysis. 

Neural Networks (NN) are composite non-linear functions with a large number of parameters\footnote{An example is the GPT-3 model, which has 175 billion parameters, as detailed in \cite{brown2020language}.}. Training an NN for a supervised learning (SL) task, much like solving a satisfiability problem, is inherently a highly non-convex optimization problem. However, there are some surprising findings that warrant further understanding: i) remarkably, simple algorithms such as gradient-based methods, which are originally designed for convex problems, can be adapted to become highly effective for learning tasks; ii) these algorithms yield solutions with good generalization properties. This is unexpected because neural networks are large systems with more free parameters than constraints and therefore  could potentially overfit. In SL one is given: i) a training set $\mathcal{D}=\{(\boldsymbol{\xi}^{\mu},y^{\mu})\}_{\mu=1}^P$, consisting of $P$ pairs, each pair being an association between a data point $\boldsymbol{\xi}^{\mu}\in \mathbb{R}^N$ and a label $y^{\mu}$; ii) a NN, that is a function $f_{\boldsymbol{W}}(\cdot)$ depending on a set of parameters $\boldsymbol{W}$ mapping the space of data into the space of labels. The goal is to find a realization of $\boldsymbol{W}$ that allows the network to fit $\mathcal{D}$. In formulas, $\boldsymbol{W}$ should satisfy:
\begin{equation}
\label{eq:training}
f_{\boldsymbol{W}}(\boldsymbol{\xi}^{\mu})=y^{\mu},\quad \forall\,\,\,\, (\boldsymbol{\xi}^{\mu},y^{\mu})\in\mathcal{D}.
\end{equation}
Given a solution $\boldsymbol{W}^*$ to \eqref{eq:training}, one is interested in its \emph{generalization properties}, which refer to how accurately $f_{\boldsymbol{W}^*}$ can predict output labels for new, unseen data points. The ultimate goal is to find solutions to \eqref{eq:training} that exhibit high generalization performance.

Problem \eqref{eq:training} can be translated into the statistical mechanics language by defining a Gibbs-Boltzmann distribution with an energy function, commonly referred to as the loss function in the context of NNs, which counts the number of errors on the training set. In the limit of zero temperature one obtains the flat distribution on all the solutions. 

The simplest NN architecture that is tractable with analytical techniques, while possessing interesting geometrical properties is the binary perceptron. In this case $\boldsymbol{W}\in \{\pm 1\}^N$, and $f_{\boldsymbol{W}}(\boldsymbol{\xi}^{\mu})=\text{sign}(\boldsymbol{W}\cdot \boldsymbol{\xi}^{\mu})$. Two data generation processes are mainly studied in the physics literature. One is the ``storage'' setting, where the $y^{\mu}$'s are i.i.d.\ Rademacher variables. The other is the ``teacher-student'' setting, where one extracts a ``teacher'' weight $\boldsymbol{W}^T$, whose entries are i.i.d. Rademacher, and then defines the labels as $y^{\mu}=\text{sign}(\boldsymbol{W}^T\cdot \boldsymbol{\xi}^{\mu})$. In both cases the $\boldsymbol{\xi}^{\mu}$ are extracted independently with entries $\xi^{\mu}_i\sim\mathcal{N}(0,1/N)$. The flat measure on all the solutions is defined by the partition function:
\begin{equation}
Z=\sum_{\boldsymbol{W}}\mathbb{X}_{\mathcal{D}}(\boldsymbol{W}),\quad \mathbb{X}_{\mathcal{D}}(\boldsymbol{W})=\prod_{\mu=1}^P\Theta\left(y^{\mu}\,\boldsymbol{W}\cdot\boldsymbol{\xi}^{\mu}\right).
\end{equation}
In the limit of $N,P\rightarrow \infty$ with fixed ratio $\alpha=P/N$ it is found that: i) in the storage scenario, there exists a critical ``capacity'' $\alpha_c\approx 0.833$ s.t.\ the typical instance has solution with probability one for $\alpha<\alpha_{c}$ and zero otherwise; ii) in the teacher-student scenario, there exists a critical $\alpha_{ts}\approx 1.25$ s.t. for $\alpha<\alpha_{ts}$ there is an exponential number of solutions, while for $\alpha\geq \alpha_{ts}$ the solution is unique (the teacher). 

\begin{figure}
    \centering
    \includegraphics[width=0.8\textwidth]{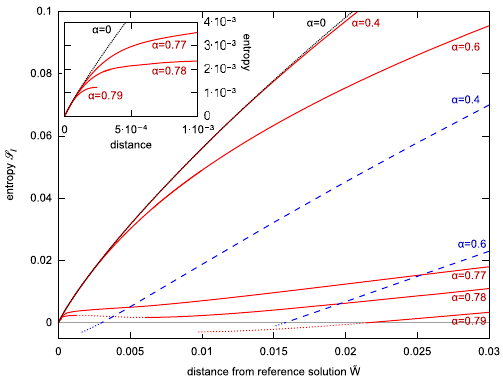}
    \caption{Local entropy $\mathcal{S}_I$ as a function of the distance from the reference solution $\boldsymbol{\tilde{W}}$. Continuous lines correspond to $y$ optimized in such a way that $\mathcal{S}_E(d,y^*)=0$ for different values of $\alpha$. The inset is a zoom on the small distances region. For optimized $y$ the local entropy is always positive for small $d$. Dashed lines represent the local entropy of a typical reference solution. For a typical reference solution, no other solutions exist at sufficiently small positive normalized Hamming distances. The dotted blue segments with negative entropy are unphysical continuations of the RS calculation. The figure is taken from \cite{baldassi2015subdominant}.}
    \label{fig:local_entropy}
\end{figure}

Studying the geometry of the space of configurations, it is found that typical solutions according to the flat measure are isolated and far-apart (golf course scenario), namely their mutual Hamming distance is $O(N)$ \cite{huang2014origin}. Additionally, there is an exponential number of local minima of the energy \cite{krauth1989storage,sompolinsky1990learning,huang2013entropy,obuchi2009weight}. These ``glassy'' states trap the dynamics of Monte Carlo algorithms satisfying detailed balance w.r.t.\ the Gibbs-Boltzmann measure. Moreover problems where \emph{all} pairs of solutions are isolated and far apart represent a special case of a disconnectivity property known as ``overlap-gap'' property (OGP) (see \cite{gamarnik2021overlap}). This property is linked with the failure of a broad class of algorithms \cite{gamarnik2018finding,gamarnik2021overlap,gamarnik2022algorithms}. So why is it that there are algorithms capable of learning efficiently in these systems? 
A large deviation analysis shows that beside the typical point-like solutions, there is another class of solutions belonging to entropic clusters that violate OGP. The latter are subdominant according to the flat measure. It turns out that the solutions most relevant to algorithms are not the typical ones, but rather those belonging to clusters. Two main reasons make these ``entropic'' solutions interesting: i) they attract a large class of local distributed algorithms \cite{braunstein2006learning,baldassi2015max,baldassi2007efficient,baldassi2009generalization} ; ii) they generalize well. In order to study the entropic clusters, \cite{baldassi2015subdominant} introduced a large-deviation free energy, statistically favoring solutions that are surrounded by a large number of other solutions:
\begin{equation}
\label{eq:reweightedFreeEntropy}
\mathcal{F}(d,y)=-\frac{1}{Ny}\mathds{E}_{\mathcal{D}}\log {\sum_{\boldsymbol{\tilde{W}}}\mathbb{X}_{\mathcal{D}}(\boldsymbol{\tilde{W}})\,\mathcal{N}_\mathcal{D}(\boldsymbol{\tilde{W}},d)^y},
\end{equation}
where 
\begin{equation}
\mathcal{N}_{\mathcal{D}}(\boldsymbol{\tilde{W}})=\sum_{\boldsymbol{W}}\mathbb{X}_{\mathcal{D}}(\boldsymbol{W})\,\delta\left(\boldsymbol{W}\cdot\boldsymbol{\tilde{W}},N(1-2d)\right)
\end{equation}
counts the number of solutions having a given Hamming distance $dN$ from a reference solution $\boldsymbol{\tilde{W}}$. The parameter $y$, not to be confused with the labels $y^{\mu}$, plays the role of an inverse temperature. In particular, higher values of $y$ give greater weight to more entropic solutions. An important quantity allowing to test the existence of these rare clusters of solutions is the local entropy $\mathcal{S}_I(d,y)=-\partial_y(y\mathcal{F}(d,y))$, also called internal entropy. Indeed entropic solutions are characterized by $\mathcal{S}_I(d,y)>0$ in a neighborhood of $d=0$. Another interesting quantity is the entropy of the reference configurations $\boldsymbol{\tilde{W}}$, $\mathcal{S}_E(d,y)=-y\mathcal{F}(d,y)-y\mathcal{S}_I(d,y)$, also called external entropy. In \cite{baldassi2015subdominant}, the free energy \eqref{eq:reweightedFreeEntropy} is computed with the replica method at the RS level. It is found that for each given $\alpha$ and $d$, there is a limiting value of $y$, $y^*(d,\alpha)$, such that $\mathcal{S}_E(d,y^*)=0$. For $y>y^*(d,\alpha)$ the RS solution is inconsistent, since the entropy cannot be negative in a system with binary variables. The authors of \cite{baldassi2015subdominant} choose for each $\alpha,d$ the maximum value $y=y^*(d,\alpha)$ for which RS is consistent, in order to focus on the most entropic clusters. Note that the condition $\mathcal{S}_E(d,y^*)=0$, implies that the ``optimal'' reference configurations, namely the one maximizing the local entropy, are at most sub-exponential. The study of the local entropy unveils a very rich geometry of the configuration space (see Fig.~\ref{fig:local_entropy}) that is qualitatively similar for both the storage and the teacher-student cases. It is shown that for all $\alpha$ smaller than $\alpha_c$ in the storage and $\alpha_{ts}$ in the teacher student, there is always a region around $d=0$ in which the local entropy is positive. However there is a critical value $\alpha_U$, where $\alpha_U\approx 0.77<\alpha_c$ in the storage case and $\alpha_U\approx 1.1<\alpha_{ts}$ in the teacher-student, such that for $\alpha>\alpha_U$ the RS solution is inconsistent. This inconsistency is related with the fact that at $\alpha_U$ the local entropy starts to be a non-monotonic function of $d$. It is conjectured that: i) below $\alpha_U$ the densest clusters of solutions are all connected, while above $\alpha_U$ they fracture in disconnected components; ii) the fracture of the densest clusters introduces OGP, implying that $\alpha_U$ is an algorithmic threshold, above which learning becomes hard also for smart algorithms, even if an exponential number of solutions still exists. These geometric arguments have also motivated potential cryptographic applications of neural networks. The idea is to exploit the computational barriers associated with the overlap gap property to construct functions that are easy to evaluate, while certain associated search problems are conjectured to be computationally hard \cite{benedetti2026collision,benedetti2026singlelayer,benedetti2025overlap}.

In \cite{baldassi2021unveiling} it is found that rare entropic structures that are accessible by algorithms arise as an aggregation of smaller entropic structures rooted on ``high-margin'' solutions. A solution $\boldsymbol{W}$ has margin $\kappa>0$ if it not only fits the dataset (condition \eqref{eq:training}), but also if it satisfies:
\begin{equation}
y^{\mu}\,\boldsymbol{W}\cdot \boldsymbol{\xi}^{\mu}\geq \kappa,\quad \forall\,\,\,\, (\boldsymbol{\xi}^{\mu},y^{\mu})\in\mathcal{D}. 
\end{equation}
Interestingly, this complex of subclusters exhibits a hierarchical organization. High-margin solutions, which are the ``roots'' of subclusters, are less frequent and closer to each other compared to typical solutions. These high-margin solutions are typically surrounded by other solutions with progressively lower margins, forming a nested structure. Note that high-margin imply ``robustness'' w.r.t.\ the local addition of noise to the input and the weights. A link between robustness and generalization can be made observing that solutions having good generalization properties should achieve compatible performances on an example $\boldsymbol{\xi}^{\mu}$ belonging to the training set, and on a testing sample $\tilde{\boldsymbol{\xi}}$ that is close to $\boldsymbol{\xi}^{\mu}$ \cite{xu2012robustness}. 

Interestingly, similar to survey propagation, the geometrical picture emerging from the replica analysis inspires the development of new algorithms here as well. These algorithms perform exceptionally well and are widely used in the machine learning community \cite{baldassi2016unreasonable,chaudhari2019entropy,foret2020sharpness}.

\subsubsection*{Overparametrization}

The phenomenon of overparameterized neural networks (NNs) performing well \cite{neyshabur2014search}, despite having more parameters than training data, is entirely unforeseen given the traditional intuition of the bias-variance trade-off. Indeed as model complexity increases, one would expect the error introduced by approximating the ground truth by a simplified model (bias) to decrease, and the error introduced by the model’s sensitivity to small fluctuations in the training set (variance) to increase, ultimately leading to overfitting. This problem has been largely studied by computer scientists and in the statistical physics community \cite{goldt2020modeling,d2020double,gerace2020generalisation,rocks2022memorizing}. In this section, we discuss some of the results from \cite{baldassi2022learning}, where a simple overparametrized NN is examined from the geometric large deviation perspective introduced in the previous section. As a model for overparametrization, \cite{baldassi2022learning} use a two-layers random feature model (RFM), in which the weights of the first layer are independent random variables (random features), encoded in a $N\times D$ matrix $\boldsymbol{F}$ with i.i.d.\ entries $F_{ik}\sim\mathcal{N}(0,1)$, and the weights of the second layer are binary variables $\boldsymbol{W}\in\{\pm 1\}^N$. In formulas, given an input $\boldsymbol{\xi}^{\mu}\in \mathbb{R}^D$, the network produces an output label $y^{\mu}$ as follows:
\begin{equation}
\label{eq:RFM}
y^{\mu}=\text{sign}\left(\frac{1}{\sqrt{N}}\sum_{i=1}^NW_i\,\text{sign}\left(\frac{1}{\sqrt{D}}\sum_{k=1}^DF_{ik}\xi^{\mu}_k\right)\right).
\end{equation}
The NN is asked to fit a dataset $\mathcal{D}=\{(\boldsymbol{\xi}^{\mu},y^{\mu})\}_{\mu=1}^P$ generated by a binary ``teacher'' perceptron:
\begin{equation}
y^{\mu}=\text{sign}\left(\frac{1}{\sqrt{D}}\sum_{k=1}^DW^T_k\xi^{\mu}_k\right),
\end{equation}
where $\boldsymbol{W}^T\in\{\pm 1\}^D$ have i.i.d.\ Rademacher entries, and the $\boldsymbol{\xi}^{\mu}$'s are extracted independently with entries $\xi^{\mu}_i\sim\mathcal{N}(0,1)$. Here the learning process involves only $\boldsymbol{W}$, the ``student'', while the random features are extracted at the beginning, and kept fixed afterwards. The feature layer of \eqref{eq:RFM} can be seen as a projection from a ``latent'' space (hidden manifold) where the data are generated, to a ``visible'' space that is accessible to the student $\boldsymbol{W}$. When $N>D$ this construction mimics the redundancy present in real world data, in which the dimension of the latent space is typically much smaller than that of the visible one \cite{goldt2020modeling}. 
In the limit $N,P,D\rightarrow\infty$, keeping fixed the ratios:
\begin{equation}
\alpha=\frac{P}{N},\quad \alpha_T=\frac{P}{D},\quad \alpha_D=\frac{D}{N},
\end{equation}
the system displays several interesting geometrical features. These are studied in \cite{baldassi2022learning} with a local entropy formalism that is slightly simplified with respect to the one presented before, and is based on the study of the local entropy properties of solutions with margin.
\begin{figure}
    \centering
    \includegraphics[width=0.8\textwidth]{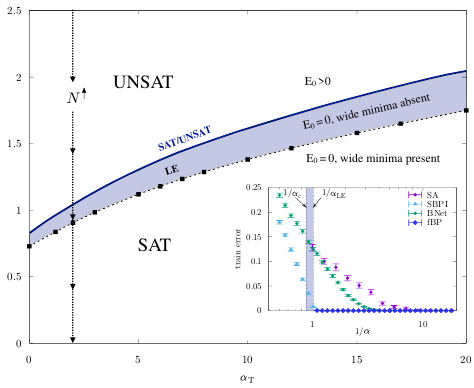}
    \caption{Phase diagram of the teacher-student RFM. At fixed $P$, for small values of $N$ (large $\alpha$) the problem is UNSAT. Increasing $N$, fitting the dataset becomes information-theoretically possible at the interpolation transition (continuous line). The light-blue region is conjectured to be algorithmically hard: solutions exist, but are hard to find. Below the dashed line (local-entropy transition line) a large connected entropy cluster appears. \emph{Inset}: training error obtained with different learning algorithms as a function of the degree of overparametrization $1/\alpha$ for fixed $\alpha_T$. None of the tested algorithms is able to fit the dataset for $\alpha>\alpha_{LE}$. The figure is taken from \cite{baldassi2022learning}.}
    \label{fig:RFM_overp}
\end{figure}
Three phases are identified in the plane $(\alpha_T,\alpha)$ (see Fig.~\ref{fig:RFM_overp}). For a given $\alpha_T$, there are two critical values $\alpha_c(\alpha_T)$ and $\alpha_{LE}(\alpha_T)$, with $\alpha_{LE}(\alpha_T)<\alpha_c(\alpha_T)$ s.t.\ : i) for $\alpha>\alpha_c(\alpha_T)$, the so-called ``interpolation'' transition, the problem is UNSAT, namely for a typical instance it does not exist a student able to fit the dataset; ii) for $\alpha_{LE}(\alpha_T)<\alpha<\alpha_c(\alpha_T)$ the student has a large enough number of parameters to be able to learn, namely solutions exist in this region. However typical solutions are isolated, and atypical entropic solutions belong to disconnected clusters. This region is conjectured to be algorithmically hard; iii) for $\alpha<\alpha_{LE}(\alpha_T)$, the ``local entropy'' transition, the dense entropic clusters rooted on high-margin solution coalesce. It is found that in this region algorithms are able to learn. Therefore if the model is enough overparametrized, learning becomes easy due to the appearance of a rare connected structure with high local entropy. It is important to keep in mind that even in the easy phase there is an exponential number of poor-generalizing isolated and point-like solutions dominating the flat measure, and algorithms targeting these solutions take an exponential time in $N$ to succeed.

\subsubsection*{Asymmetric Hopfield Model}

The modeling of associative memories is another example of research lines that significantly benefited from statistical physics techniques. A paradigmatic example is the Hopfield model (HM) \cite{hopfield1982neural}. The HM is a model of associative memory, where a specific learning rule allows the system to be designed such that the minima of the free energy landscape correspond to the desired memory configurations. In its simplest form, the HM resembles the SK model, as both are defined by the same energy function (see Eq.~\eqref{eq:spinglassdef}). However the difference is that the couplings, that in SK are i.i.d.\ random variables, in the HM are determined by the learning rule and are related to the stored memories. A significant limitation of the HM is that it relies on symmetric couplings, $J_{ij}=J_{ji}$, which poses several problems on the performance of the memory \cite{parisi1986asymmetric}, and is highly unrealistic from a physiological point of view. Indeed in biological systems the interactions between neurons are not necessary reciprocal. For this reason, great efforts have been put in the study of neural networks with asymmetric couplings \cite{sompolinsky1988chaos}. The presence of asymmetry implies that the dynamics of the system becomes chaotic. For example we could have that the state of a spin becomes a periodic function of time, with period that could be also exponentially large in the size of the system. 

In this section we discuss some geometrical properties of the configuration space of one of such networks with asymmetric couplings \cite{badalotti2026dynamical}. Consider the dynamics:
\begin{equation}
\label{eq:dynHasym}
s_i^{t+1}=\text{sign}\left(J_Ds_i^t+\sum_{j\neq i}J_{ij}s_j^{t}\right),
\end{equation}
where the $J_{ij}$'s are random variables with $P(J_{ij},J_{ji})=P(J_{ij})P(J_{ji})$ and $P(J_{ij}=\pm 1/\sqrt{N})=1/2$. The term $J_D s_i$ is a self-interaction, where $J_D$ is a control parameter that determines an interplay between chaos and ``bistability'' \cite{stern2014dynamics}: for $J_{D}$ large enough the spins (neurons) decouple, and every configuration becomes a fixed point; for $J_{D}=0$ the number of fixed points is at most sub-exponential in $N$ (it can be shown that the entropy of the fixed points vanishes at $J_D=0$). 

The study of the fixed points of \eqref{eq:dynHasym} is crucial, as these fixed points correspond to memories when the couplings are chosen according to a properly-defined learning rule. Here we discuss some of the outcomes of this study in a simplified setting, where the $J_{ij}$'s are random. 

Using the optimization language introduced before, it is possible to define an ``energy'' function that is zero if a configuration satisfies the fixed point condition deriving from \eqref{eq:dynHasym} and greater than zero otherwise: 
\begin{equation}
E_J(\boldsymbol{s})=\sum_i\left[1-s_i\text{sign}\left(J_Ds_i+\sum_{j\neq i}J_{ij}s_j\right)\right].
\end{equation}
Now consider the partition function at inverse temperature $\beta$: 
\begin{equation}
Z_J=e^{-\beta N}\sum_{\boldsymbol{s}}e^{\beta \sum_i s_i\text{sign}\left(J_Ds_i+\sum_{j\neq i}J_{ij}s_j\right) }.
\end{equation}
It turns out that $Z_J$ is self-averaging \cite{badalotti2026dynamical}, and therefore the free energy density $f$ can be found by an annealed computation:
\begin{equation}
\label{eq:freeEnHM}
f=-\lim_{N\rightarrow \infty}\frac{1}{\beta N}\log\mathds{E}_J Z_J=1-\frac{1}{\beta}\left[\log{2}+\log{\big(e^{\beta}H(-J_D)+e^{-\beta}H(J_D)\big)}\right],
\end{equation}
where 
\begin{equation}
H(x)=\int_x^{\infty}\frac{dh}{\sqrt{2\pi}}e^{-\frac{h^2}{2}}.
\end{equation}
In the limit $\beta\rightarrow \infty$, Eq.~\eqref{eq:freeEnHM} allows to compute the entropy density $s$ of the fixed points:
\begin{equation}
\label{eq:EntroHM}
s=-\lim_{\beta\rightarrow \infty}\beta\,f=\log{2}+\log{(1-H(J_D))}.
\end{equation}
Note that $s(J_D=0)=0$. Equation \eqref{eq:EntroHM} implies that for $J_D>0$ there is an exponential number of fixed points of the dynamics \eqref{eq:dynHasym}. However, these typical fixed points are isolated and unstable, and the dynamics of the network is in general chaotic. The question is if there are other states that are perhaps not dominant but stable. A local-entropy analysis of the problem reveals the following picture \cite{badalotti2026dynamical}: i) for $0< J_D \leq J_c\approx 0.1$ typical unstable solutions coexist with an exponential number of isolated dense clusters of stable solutions that are subdominant according to the flat measure. The naive dynamics \eqref{eq:dynHasym} is chaotic, and stable fixed points are conjectured to be hard to find; clusters of solutions satisfy the overlap gap property and are expected to be impossible to access with local algorithms; ii) for $J_c<J_D<J_{alg}\approx 0.8$ (the ``liquid'' phase) the atypical dense clusters coexist with an exponential number of regions that connect them. Despite the naive dynamics being chaotic, atypical stable solutions can be found with efficient algorithms; iii) for $J_D>J_{alg}$, the naive dynamics also finds stable fixed points.

As for the learning problem, the existence of rare dense regions is crucial for algorithmic accessibility.

\section{Conclusions}

We have presented concrete examples of discrete optimization problems and neural network models in which the geometry associated with RSB helps explain computational barriers and algorithmic accessibility.

For continuous models, the picture can become even more involved, with full RSB effects playing a crucial role.

These examples highlight the continuing influence of Parisi’s ideas on computer science and machine learning. We hope this chapter encourages reflection on the development of RSB theory and on its future applications in these intertwined disciplines.

\newpage
\begingroup

\bibliographystyle{unsrturl}
\bibliography{bibliography.bib}

\endgroup

\end{document}